\documentclass[doublecol]{epl2}

\usepackage{graphicx}
\usepackage{bm}
\usepackage{color}
\usepackage{amssymb}

\graphicspath{{./plots/}}

\newcommand{\lla}{\left\langle}
\newcommand{\rra}{\right\rangle}

\newcommand{\wi}{\mathrm{Wi}}
\newcommand{\wic}{\mathrm{Wi}_c}

\title{Tumbling of polymers in semidilute solution under shear flow}
\shorttitle{Tumbling Dynamics of Polymer in Semidilute Solution} %Insert here a short version of the title if it exceeds 70 characters

\author{C.-C. Huang\inst{1} \and G. Sutmann\inst{2}
\and G. Gompper\inst{1,} \inst{3} \and R. G. Winkler\inst{3} }
\shortauthor{C.-C. Huang \etal}

\institute{
  \inst{1} %
Institute of Complex Systems, Forschungszentrum J\"{u}lich,
D-52425 J\"{u}lich, Germany\\
  \inst{2} J\"{u}lich Supercomputing Centre, Forschungszentrum J\"{u}lich,
D-52425 J\"{u}lich, Germany\\
  \inst{3} Institute for Advanced Simulation, Forschungszentrum J\"{u}lich,
D-52425 J\"{u}lich, Germany
}

\abstract{
The tumbling dynamics of individual polymers  in semidilute
solution is studied by large-scale non-equilibrium mesoscale hydrodynamic
simulations. We find that the tumbling time is equal to the non-equilibrium
relaxation time of the polymer end-to-end distance along the flow
direction and strongly depends on concentration. In addition, the normalized
tumbling frequency as well as the widths of the alignment distribution
functions for a given concentration dependent Weissenberg number
exhibit a weak concentration
dependence in the cross-over regime from a dilute to a semidilute solution.
For semidilute solutions a universal behavior is obtained.
This is a consequence of screening of hydrodynamic interactions at polymer
concentrations exceeding the overlap concentration.
}

\begin{document}

\maketitle
\section{Introduction}
Polymers in solution exposed to shear flow exhibit a remarkably
rich dynamical behavior, as has been shown by direct observation
using fluorescence
microscopy~\cite{smith:99,schr:05,teix:05,schr:05_1,gera:06}. In
particular, polymers exhibit tumbling motion, i.e., they undergo a
cyclic stretch and collapse dynamics, with a characteristic
frequency which depends on the shear rate and their internal
relaxation time. This nonequilibrium behavior has intensively been
studied for polymers in dilute solution
~\cite{smith:99,aust:99,schr:05,teix:05,gera:06,doyl:00,cela:05,cher:05,puli:05,schr:05_1,delg:06,wink:06_1,zhan:09}.

The dynamical behavior of a polymer in \textit{semidilute}
solution under shear flow has received far less attention
\cite{babc:00,hur:01,huan:10_1}. Insight into the behavior of such
systems is of fundamental importance in a wide spectrum of systems
ranging from biological cells, where transport appears in dense
environments, to turbulent drag reduction in fluid flow. While the
dynamical behavior of polymers in dilute solution is strongly
affected by hydrodynamic
interactions~\cite{doi:86,kapr:08,gomp:09}, their relevance in
semidilute solutions is less clear.
%
%
%Moreover, in semidilute solutions of long polymers, viscoelastic
%effects play an important role. Due to the long structural
%relaxation time, the internal degrees of freedom of a polymer
%cannot relax sufficiently fast under non-equilibrium conditions
%and an elastic restoring force tries to push the system towards
%its original state.
%

The complex interactions in semidilute solutions hamper an
analytical treatment. Here, computer simulations are an important
tool to shed light on the rich and intricate dynamical behavior of
such systems. The large length- and time-scale gap between the
solvent and macromolecular degrees of freedom requires a mesoscale
simulation approach in order to assess their structural,
dynamical, and rheological properties. We apply a hybrid
simulation approach, combining molecular dynamics simulations (MD)
for the polymers with the multiparticle collision dynamics (MPC)
method describing the solvent~\cite{male:99,kapr:08,gomp:09}.

By this approach, we demonstrated that polymers in dilute and
semidilute solutions exhibit large deformations and a strong
alignment along the flow direction in simple shear flow
\cite{huan:10_1}. More importantly, in the stationary state, the
{\em conformational} and {\em rheological properties} for various
concentrations are universal functions of the Weissenberg number
$\wic = \dot \gamma \tau(c)$, where $\dot \gamma$ is the shear
rate and $\tau(c)$ the concentration-dependent polymer end-to-end
vector relaxation time at equilibrium. Hence, with increasing
concentration, hydrodynamic interactions affect the conformational
and rheological properties only via the increasing relaxation time
$\tau(c)$. Experiments on DNA in shear flow \cite{hur:01} and
simulations of polymer brushes \cite{galu:10} lead to a similar
conclusion. Then, the question arises to what extent hydrodynamic
interactions are relevant in non-equilibrium systems.

In this letter, results are presented for the concentration
dependence of the non-equilibrium {\em dynamical properties} of
polymers in shear flow by calculating tumbling times and
orientational distribution functions. These quantities exhibit a
dependence on hydrodynamic interactions in dilute solution, and
are independent of such interactions in semidilute solution where
hydrodynamic interactions are screened. This is supported by a
comparison of non-draining and free-draining simulations. As a
result, hydrodynamic interactions ar found to clearly contribute
to the non-equilibrium polymer dynamics in dilute solution beyond
the change of relaxation times.

\section{Model and Parameters}

A solution is considered of $N_p$ linear flexible polymer chains
embedded in an explicit solvent. Each polymer is comprised of
$N_m$ beads of mass $M$, which are connected by linear springs of
equilibrium bond length $l$ \cite{gomp:09,huan:10_1}. Inter- and
intramolecular excluded-volume interactions are taken into account
by the repulsive, shifted and truncated Lennard-Jones potential,
with the parameter $\sigma$ characterizing the bead size and
$\epsilon$ the energy \cite{huan:10_1}. The monomer dynamics is
determined by Newton's equations of motion, which are integrated
by the velocity Verlet algorithm with time step $h_p$
\cite{alle:87}.

The solvent is simulated by the multiparticle collision dynamics
(MPC) method \cite{male:99,kapr:08,gomp:09}. It is composed of
$N_s$ point-like particles of mass $m$.  The algorithm consists of
alternating streaming and collision steps. In the streaming step,
the solvent particles move ballistically for a time  $h$. In the
collision steps, particles are sorted into cubic cells of side
length $a$ and their relative velocities, with respect to the
center-of-mass velocity of their cell, are rotated around a
randomly oriented axis by a fixed angle $\alpha$.

The solvent-polymer coupling is achieved by taking the monomers
into account in the collision step. To insure Galilean invariance,
a random shift is performed at every collision
step~\cite{ihle:01}. The collision step is a stochastic process,
where mass, momentum and energy are conserved, which leads to the
build-up of correlations between the particles and gives rise to
hydrodynamic interactions \cite{gomp:09}.

Three-dimensional periodic boundary conditions are considered for
the simulation of shorter chains. Here, Lees-Edwards boundary
conditions are applied to impose a shear flow \cite{alle:87}. A
local Maxwellian thermostat is used to maintain the temperature of
the fluid at the desired value~\cite{huan:10}. A parallel MPC
algorithm is exploited for systems of longer chains, which is
based on a three-dimensional domain-decomposition approach
\cite{huan:10_1}. In such a system, shear flow is imposed by the
opposite movement of two confining walls, and periodic boundary
conditions are applied parallel to them.  We impose no-slip
boundary conditions at walls for both, fluid particles and
monomers~\cite{lamu:01,wink:09,gomp:09}.

Non-hydrodynamic simulations are performed by Brownian MPC, where
each monomer independently performs stochastic collisions with a
phantom particle which mimics a fluid element of size $a^3$. In
shear flow, the phantom-particle momentum is taken from a
Maxwell-Boltzmann distribution with mean $ \langle p_x \rangle = m
\left\langle N_c \right\rangle \dot \gamma r_y$ and variance
$\langle p^2_{\beta}\rangle = m \left\langle N_c \right\rangle
k_BT$ ($\beta \in \{x,y,z\}$), where $\left\langle N_c
\right\rangle$ is the average number of solvent particles per
collision cell, $r_y$ the particle position along the gradient
direction, and $\langle p_x \rangle$ the momentum along the flow
direction \cite{ripo:07,gomp:09}.

We employ the parameters $\alpha = 130^{\circ}$, $h= 0.1 \tilde
\tau$, with $\tilde \tau = \sqrt{ma^2/(k_BT)}$ ($k_B$  is
Boltzmann's constant and T is temperature), $\left\langle N_c
\right\rangle = 10$, $M=m\left\langle N_c \right\rangle$,
$l=\sigma =a$, $k_BT/\epsilon =1$,  $h/h_p = 50$, and the bond
spring constant $\kappa =5\times10^3 k_BT/a^2$. The polymer
lengths $N_m=50$ and $250$ are considered in the concentration
ranges $c/c^* = 0.16 - 2.08$ and $0.17 - 10.38$, respectively. The
corresponding overlap concentrations are $c^* =0.098 l^3$ and
$c^*=0.029 l^3$, determined by their radii of gyration. In dilute
solution, the equilibrium end-to-end vector relaxation times are
$\tau_0 = 6169 \tilde \tau$ and $78330 \tilde \tau$
\cite{huan:10_1}. The Brownian MPC simulations for $N_m=50$ yield
an approximately five times larger relaxation time than
hydrodynamic MPC.

\section{Tumbling Dynamics}
\begin{figure}[t]
\begin{center}
  \includegraphics*[width=.23\textwidth,angle=270]{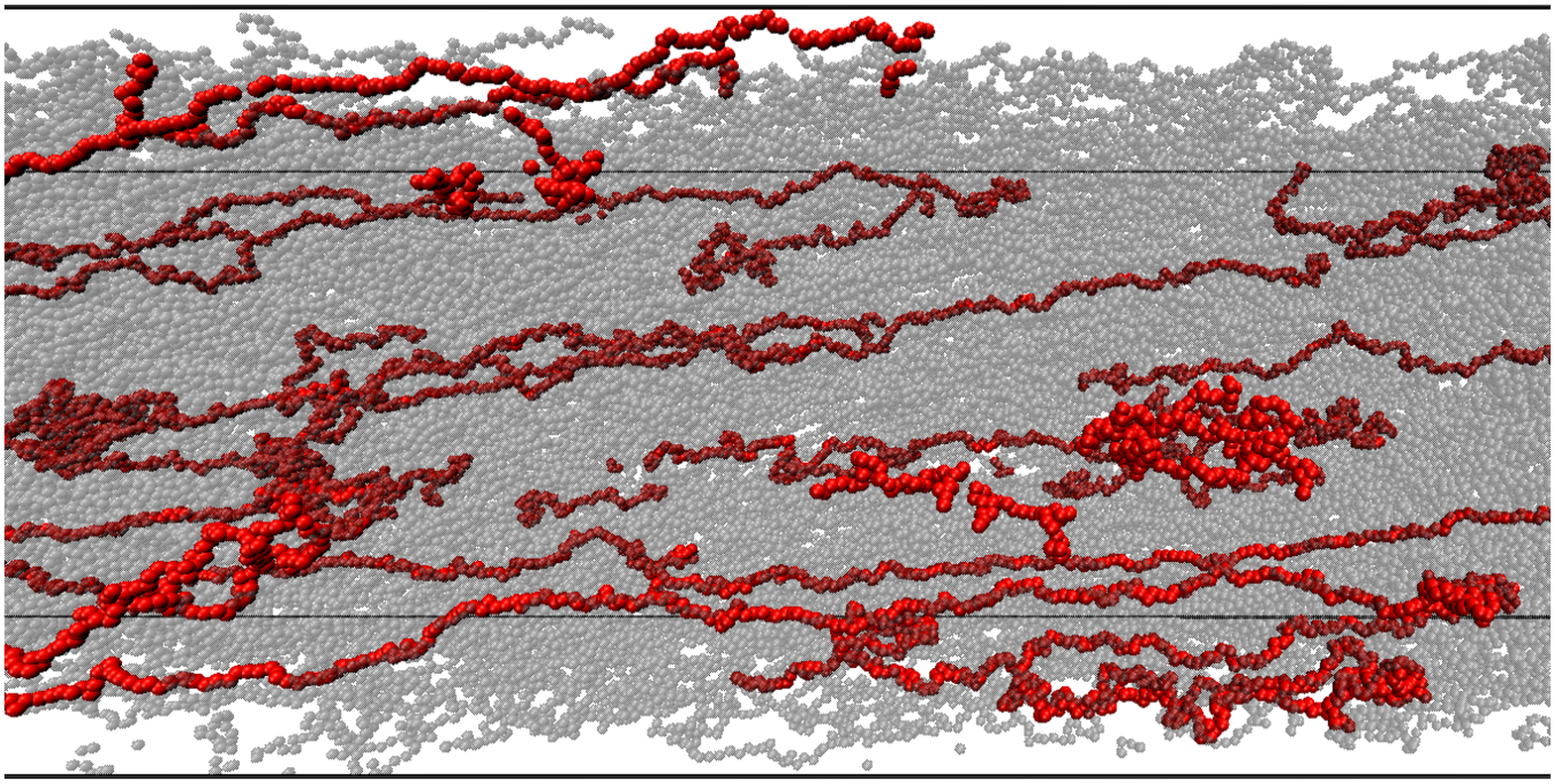}
\caption{Snapshot of a system of $N_p=800$ polymers of length $N_m=250$ for the
Weissenberg number $\wic=184$ at the concentration $c/c^* = 2.77$.
For illustration, some chains are highlighted in red. Animations are provided as
supplementary material. low.mov shows the polymer dynamics
under shear for the Weissenberg number $\wic = 18$ and heigh.mov for $\wic = 184$.}
\label{snap}
\end{center}
\end{figure}

\begin{figure}
\begin{center}
  \includegraphics*[width=8cm]{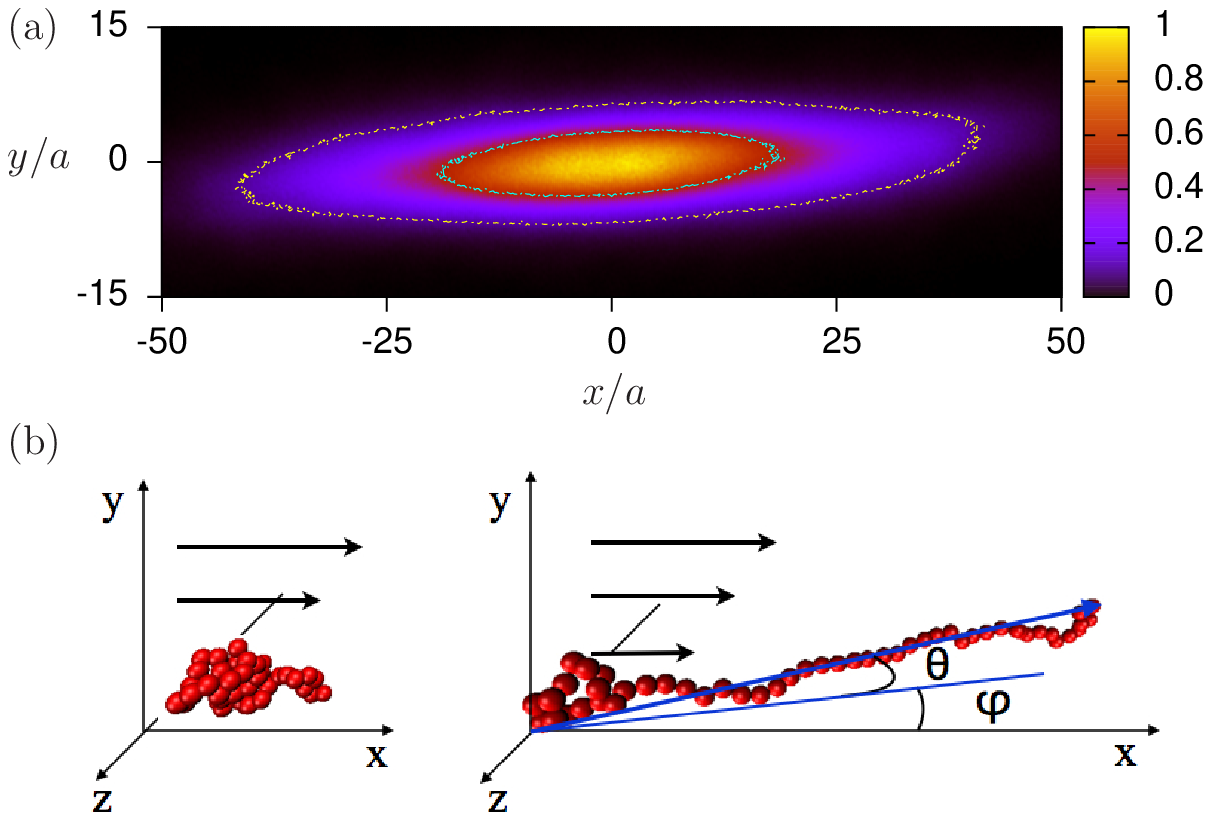}\\[1ex]
  \caption{(a) Monomer density distribution in the
  flow-gradient plane for $N_p=3000$, $N_m=250$, i.e., $c/c^*=10.38$, and
  $\wic=569$.
 The contour lines for the densities $0.1$ (outer) and $0.5$ (inner) are highlighted
  to emphasize the non-ellipticity of the shape.
  (b) Illustration of polymer stretching (right)
  and recoiling (left).
  $\theta$ is the angle between the end-to-end vector and
  its projection onto the flow-gradient plane and $\varphi$ is the angle
  between this projection and the flow direction. }
  \label{den_illu}
\end{center}
\end{figure}
%\begin{figure}
%\begin{center}
%  \includegraphics*[width=0.15\textwidth,angle=270]{figure_2a.eps}\\[1ex]
%  \includegraphics*[width=0.4\textwidth]{figure_2b.eps}
%  \caption{(a) Monomer density distribution in the
%  flow-gradient plane for $N_p=3000$, $N_m=250$, i.e., $c/c^*=10.38$, and
%  $\wic=569$.
% The contour lines for the densities $0.1$(outer) and $0.5$(inner) are highlighted
%  to emphasize the non-ellipticity of the shape.
%  (b) Illustration of polymer stretching (right)
%  and recoiling (left).
%  $\theta$ is the angle between the end-to-end vector and
%  its projection onto the flow-gradient plane and $\varphi$ is the angle
%  between this projection and the flow direction. }
%  \label{den_illu}
%\end{center}
%\end{figure}

The snapshot of a semidilute solution, displayed in
Fig.~\ref{snap}, indicates large conformational differences
between the various polymers in flow. The average shape of an
individual chain is illustrated in Fig.~\ref{den_illu}(a) by the
monomer density distribution in the flow-gradient plane. Their
conformational and rheological properties are discussed in detail
in Ref. \cite{huan:10_1}. The large conformational variations are
due to a continuous end-over-end tumbling motion
\cite{smith:99,schr:05}, with stretched and coiled states as
depicted in Fig.~\ref{den_illu}(b).

\begin{figure}[t]
%\vspace{0.6cm}
\begin{center}
\includegraphics*[width=.38\textwidth]{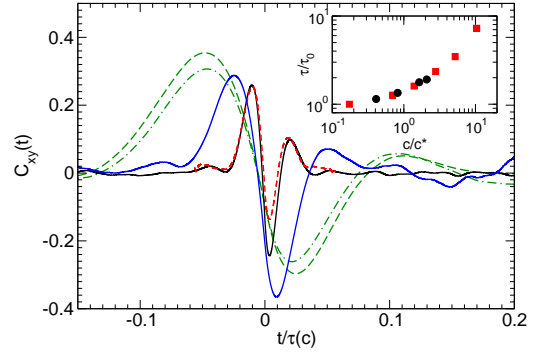}
\caption{Cross-correlation functions  [Eq. (\ref{correlation})]
for a polymer of length $N_m=250$ and the concentrations $c/c^*=2.77$
(black solid line), $c/c^*=10.38$, (red dashed line),
and $c/c^*=0.35$ (blue solid line),
corresponding to the Weissenberg numbers $\wic=5520$, $\wic=5690$,
and $\wic=2670$, respectively, as well as $N_m=50$ with $c/c^*=2.08$
(green dashed-dotted line) and $c/c^*=0.8$ (green dashed line),
the Weissenberg numbers
are $\wic = 220$ and $\wic=233$, respectively.
Inset: Longest polymer relaxation times for
$N_m=50$ ($\bullet$) and $N_m=250$ (${\color{red}\blacksquare}$).} \label{cross}
\end{center}
\end{figure}

The instantaneous shape of a polymer is characterized by the
radius of gyration tensor $G_{\beta \beta'}$ ($\beta, \beta' \in
\{x,y,z\}$), which  is defined as
$G_{\beta\beta'}=\sum^{N_m}_{i=1}\langle\Delta r_{i,\beta} \Delta
r_{i,\beta'}\rangle /N_m$, where $\Delta {\bm r}_{i}$ is the
position of monomer $i$ in the center-of-mass reference frame of
the polymer.  To find a characteristic time for the tumbling
dynamics, we determine the cross-correlation function
\begin{equation} \label{correlation}
C_{xy}(t)=\frac{\lla G_{xx}'(t_0)G_{yy}'(t_0+t)\rra }
{\sqrt{ \lla G_{xx}'^{2}(t_0) \rra \lla G_{yy}'^{2}(t_0) \rra }} ,
\end{equation}
for deviations from average stationary values
$G_{\beta\beta}'(t)=G_{\beta\beta}(t)- \lla G_{\beta\beta} \rra$.
% of the radius of gyration tensor components along the flow
%($x$-axis) and gradient ($y$-axis) direction.
Figure~\ref{cross}
shows cross-correlation functions for several shear rates and
concentrations. Each of the curves exhibits a deep minimum at time
$t_{+}>0$ and a pronounced peak at time $t_{-}<0$, and decays to
zero at large time-lags. Hence, the tumbling dynamics is not
periodic, but cyclic. The latter has been questioned for tethered
polymers \cite{zhan:09}. The minimum at $t_{+}$ indicates that
positive values of $G_{\beta \beta}'$ are linked with negative
ones of the orthogonal directions, i.e., polymer shrinkage in the
$y$-direction is linked with its extension in $x$-direction, and
similarly, an extension in $y$-direction is linked to shrinkage in
$x$-direction. The maximum of $C_{xy}(t)$ reveals that positive
deviations $G_{xx}'$ are correlated with positive values $G_{yy}'$
at earlier times, or a collapsed state along the $x$-direction
($G_{xx}' <0$) is correlated with a previous collapsed state in
$y$-direction \cite{schr:05}. Hence, the time difference $t_{+}-
t_{-}$ is related to conformational changes that a polymer
undergoes during tumbling. We therefore characterize tumbling by
the time $\tau_t = 2(t_{+}-t_{-})$. The factor two is introduced,
because two non-equivalent conformations lead to a maximum and a
minimum, respectively, and will be (more or less) assumed during a
cycle.

As shown in Fig.~\ref{cross}, the positions of the peaks and
minima are rather close for equal Weissenberg numbers, when the
lag-time is scaled by the relaxation time $\tau(c)$. Hence, the
tumbling times exhibit a strong concentration dependence due to
the strong concentration dependence of the relaxation times, which
is shown in the inset of Fig. 3 \cite{huan:10_1}.

\begin{figure}[t]
%\vspace{0.6cm}
\begin{center}
\includegraphics*[width=.38\textwidth]{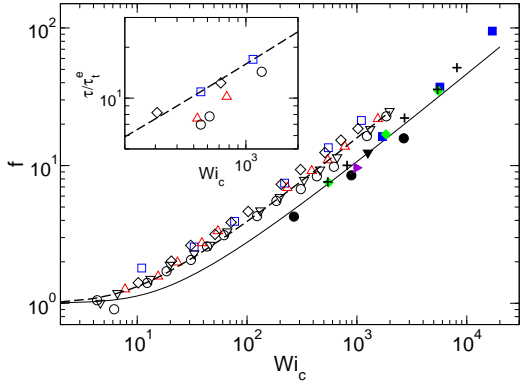}
\caption{Normalized tumbling frequencies $f=\tau(c)/\tau_t$ obtained from
cross-correlation functions.
Open symbols correspond to the polymer length
$N_m=50$ for  $c/c^*=0.16$ ($\circ$), $c/c^*=0.4$
($\triangledown$), $c/c^*=0.8$ ($\color{red}\triangle$), $c/c^*=1.6$
(${\Diamond}$), and $c/c^*=2.08$
(${\color{blue}\square}$). Filled symbols indicate results for
$N_m=250$ and $c/c^*=0.35$ ($\bullet$),
$c/c^*=0.69$ ($\blacktriangleright$), $c/c^*=1.38$
($\blacktriangledown$), $c/c^*=2.77$
(${\color{green}\blacklozenge}$), $c/c^*=5.19$
($+$), $c/c^*=10.38$
(${\color{blue}\blacksquare}$).  The lines present
the theoretical predictions \cite{wink:06_1,wink:10}.
Inset: Relaxation times $\tau^e_t$ for $N_m=50$ and various concentrations.}
\label{tumb_freq}
\end{center}
\end{figure}

Normalized tumbling frequencies $f= \tau(c)/\tau_t$, with tumbling
times extracted from the correlation functions and scaled by the
corresponding relaxation times $\tau(c)$, are presented in
Fig.~\ref{tumb_freq} for a wide range of shear rates and
concentrations. For comparison, the theoretical prediction for a
polymer in dilute solution is presented as well
\cite{wink:06_1,wink:10}. The results are in excellent agreement,
when the Weissenberg number $\wi^*$ of the theoretical model is
identified with $\wi^*=\wic/2$, where the factor two accounts for
the approximately twice larger relaxation time of the theoretical
model. The short chain results clearly show the crossover from
unity, assumed in the limit $\dot \gamma \to 0$, to the asymptotic
dependence $\sim \wic^{2/3}$ at high shear rates. We obtain a
chain-length dependence in close agreement with the theoretical
prediction. More importantly, we find a slight and gradual shift
of $f$ to larger values with increasing concentration at a given
$\wic$, until a saturation is reached in the semidilute regime
$c/c^*>1$. This is seen for the two largest concentrations for
$N_m=50$ and the three largest ones for $N_m=250$. As a
consequence, the polymers exhibit a universal behavior, both in
dilute ($c \ll c^*$) as well as in semidilute solution as function
of $\wic$, with the same power-law dependence on $\wic$ (for $\wic
>1$).

Theory~\cite{wink:06_1}, simulations~\cite{puli:05}, and
experiments~\cite{gera:06} suggest that tumbling is an aperiodic
process with an exponential distribution of  intervals between
tumbling events, i.e.,  $P_t(t)\sim \exp(-t/\tau^e_{t})$, where
$\tau^e_{t}$ is defined as the characteristic tumbling time. By
calculating  the distribution functions of times between
successive gradient-vorticity plane crossings of the end-to-end
vector, we determined the tumbling times $\tau^e_{t}$ presented in
the inset of Fig~\ref{tumb_freq}. Their dependence on Weissenberg
number and concentration is in perfect accord with that obtained
for $\tau_t$; the absolute values are only approximately $20\%$
smaller.

Alternatively, relaxation times under shear flow can be obtained
by the end-to-end vector auto-correlation function $\left\langle
R_{\beta} (t) R_{\beta}(0)\right\rangle$, where $R_{\beta} =
r_{N,\beta}-r_{1,\beta}$ \cite{wink:06_1,delg:06,jose:08}. Similar
to the results presented in Ref. \cite{jose:08}, we find a damped
oscillatory behavior for $\wic \gtrsim 1$. A fit of $g(t) =
e^{-t/\tau_{r}}[\cos(\omega t) + a \sin(\omega t)]$ to the
correlation functions along the flow and gradient directions,
yields, within the accuracy of the simulations, relaxation times
$\tau_{r} (\dot \gamma)$ and normalized tumbling frequencies $f=
\tau(c)/\tau_{r}$ equal to the values presented in
Fig.~\ref{tumb_freq}. The parameter $\omega$ of $g(t)$ is
independent of shear rate for the considered systems. This is in
contrast to results of Ref.~\cite{jose:08}, where very short
polymers ($N_m=10$) have been considered only.

The simulation results for the tumbling time lead to the following
conclusions. (i) The correlation function (\ref{correlation}) can
be used to obtain a characteristic time $\tau_t$ for the tumbling
motion (see also Ref. \cite{teix:05}). (ii) The non-equilibrium
end-to-end distance relaxation time $\tau_{r}$ along the flow
direction is equal to the tumbling time $\tau_t$ extracted from
the correlation function (\ref{correlation}). Hence, the tumbling
time $\tau_t$ is equal to the non-equilibrium relaxation time
$\tau_{r}$, as also predicted in Refs.~\cite{wink:06_1,zhan:09}.
(iii) There is a weak chain-length dependence of the tumbling
time. (iv) Screening of hydrodynamic interactions in semidilute
solutions leads to a (small) increase of the normalized tumbling
frequencies. The screening aspects are discussed in more detail in
then next section.

\section{Angular Probability Distribution Functions}

\begin{figure}[t]
\begin{center}
\includegraphics*[width=.38\textwidth]{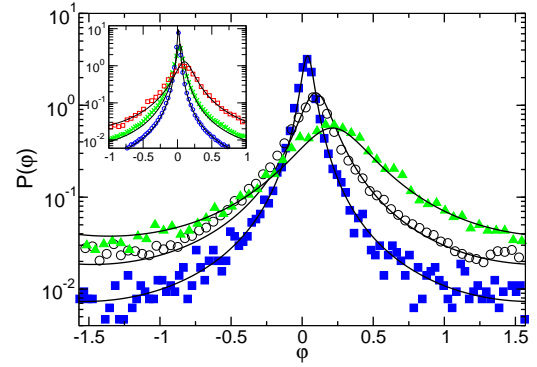}
\caption{Probability distribution functions of the angle $\varphi$ for a
dilute solution with $c/c^*=0.16$ of polymers of length $N_m=50$ for
 $\wic=616.9$
(${\color{blue}\blacksquare}$), $\wic=61.7$ ($\circ$), and $\wic=12.3$
(${\color{green}\blacktriangle}$). Inset: Distribution functions
for
$N_m=250$, $c/c^*=0.17$, and  $\wic=2350$
(${\color{blue}\circ}$), $\wic=235$ (${\color{green}\times}$),
$\wic=23.5$ (${\color{red}\square}$).
The solid lines are theoretical results for
Weissenberg numbers $\wi^*=\wic/2$.} \label{P_phi_dilute}
\end{center}
\end{figure}

Further insight into the tumbling and orientational behavior of
polymers is gained by the probability distribution functions
(PDFs) $P(\varphi)$ and $P(\theta)$ for the orientation angles
$\varphi$ and $\theta$ (cf. Fig.~\ref{den_illu}(b))
\cite{wink:06_1,puli:05,gera:06,cela:05}. For a \textit{dilute}
solution, $P(\varphi)$  is shown in Fig.~\ref{P_phi_dilute},
together with theoretical lines obtained from
Ref.~\cite{wink:06_1} (note, $\wi^* = \wic/2$). Evidently, the
simulation results  agreement well  with the analytical approach,
as is expected for a dilute solution in which the intermolecular
interactions are irrelevant. $P(\varphi)$ exhibits a significant
shear-rate dependence. At zero shear, no angle is preferred. An
increasing shear rate leads to the appearance of a peak, which
shifts to smaller values with increasing $\dot \gamma$ and, at the
same time, the width $\Delta \varphi$ of $P(\varphi)$ decreases.

\begin{figure}[t]
%\vspace{0.6cm}
\begin{center}
\includegraphics*[width=.38\textwidth]{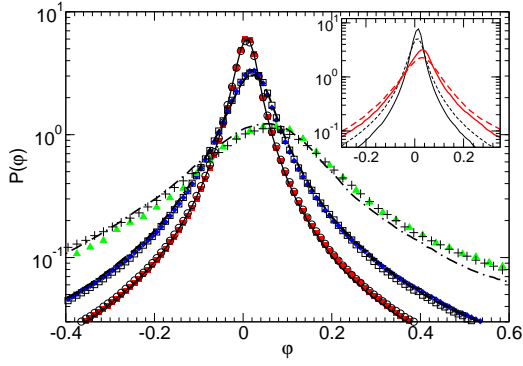}
\caption{Probability distribution functions $P(\varphi)$ of polymers of
length $N_m = 250$ for
$c/c^*=2.77$ with $\wic=5520$ (${\color{red}\blacksquare}$),
$\wic=552$ (${\color{blue}\blacklozenge}$),
and $\wic=55.2$ (${\color{green}\blacktriangle}$), for
$c/c^*=5.19$ with $\wic=5423$ ($\circ$), $\wic=542.3$ ($\square$),
and $\wic=54.23$ ($+$), as well as $c/c^*=10.38$ with $\wic=5691$ (solid line),
$\wic=569.1$ (dashed line), $\wic=56.91$ (dashed-dotted line).
Inset: Distribution functions  for $c/c^*=0.35$ and
$\wic=2670$ (black solid line), $\wic=267$ (red solid line) as well as
for $c/c^*=5.19$
with $\wic=2700$ (black dashed line) and $\wic=270$ (red
dashed line).} \label{P_phi_semidilute}
\end{center}
\end{figure}

In a \textit{semidilute} solution, the universality observed for
the tumbling time is also reflected in $P(\varphi)$.
Figure~\ref{P_phi_semidilute} displays distribution functions for
various concentrations and Weissenberg numbers $\wic$. For every
Weissenberg number distributions are compared for three
concentrations. Evidently, the distributions are independent of
concentration for the considered Weissenberg numbers.

However, we observe a clear concentration dependence, when we
compare distributions from dilute and semidilute solutions. The
inset of Fig.~\ref{P_phi_semidilute}, displays distributions for
the concentrations $c/c^*=0.35, 5.19$ and the Weissenberg numbers
$\wic \approx 267$, $2670$  and  $270$, $2700$, respectively.
Clearly, the increase in concentration from a dilute solution
beyond the overlap concentration leads to a broadening of the
distribution function.

\begin{figure}[t]
%\vspace{0.6cm}
\begin{center}
\includegraphics*[width=.38\textwidth]{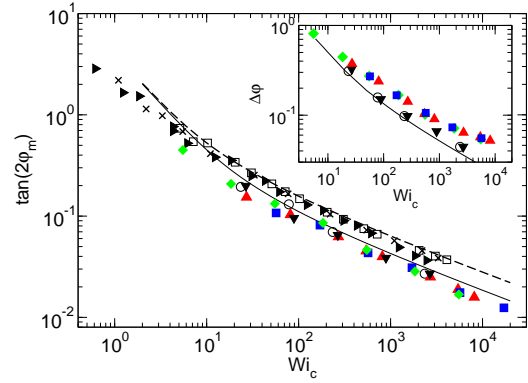}
\caption{Maximum angle $\varphi_m$ as a
function of $\wic$ for $N_m=250$ and
$c/c^*=0.17$ ($\circ$), $0.35$ ($\blacktriangledown$), $2.77$
(${\color{green}\blacklozenge}$), $5.19$
(${\color{red}\blacktriangle}$), $10.38$
(${\color{blue}\blacksquare}$), as well as $N_m=50$ for $c/c^*=0.16$
($\blacktriangleright$), $1.6$ ($\times$), and $2.08$ ($\square$).
Inset: Widths $\Delta \varphi$
of the distribution functions.
For both, the lines are theoretical predictions.}
\label{angle_e_e}
\end{center}
\end{figure}

Interestingly, the value $\varphi_m$ at the peak of the
distribution function is independent of concentration at a given
$\wic$.
%Theoretical calculations show that $\varphi_m$ is very
%close to the angle between the main axis of the radius of gyration
%tensor and the flow direction \cite{wink:06_1}.
Figure~\ref{angle_e_e} displays $\tan(2 \varphi_m)$ as function of
$\wic$ for various concentrations of the two studied chain
lengths. The simulation data are consistent with the analytical
predictions for both chain lengths \cite{wink:06_1,wink:10}. The
results show that a universal behavior is obtained for the various
concentration at a given $N_m$. In the asymptotic limit of high
shear rates, the dependence
\begin{equation}
\tan(2\varphi_m)\sim (\wic)^{-1/3}
\end{equation}
is obtained, which has also been found in Ref.~\cite{huan:10_1}
for the alignment angle determined from the gyration tensor. In
the limit of $\wic\rightarrow 0$, theory predicts
$\tan(2\varphi_m)\sim \wic^{-1}$, whereas the simulations yield
$\tan(2\varphi_m)\sim \wic^{-0.8}$ for the considered range of
$\wic$, which might be due to excluded volume interactions not
taken into account in the theoretical calculations. The inset of
Fig.~\ref{angle_e_e} displays $\Delta \varphi$, the full width at
half maximum. For dilute solutions, $\Delta \varphi$ agrees with
the prediction of the theoretical model, which yields the
asymptotic dependence $\Delta \varphi \sim \wic^{-1/3}$ in the
limit $\wic \to \infty$. The widths of the distributions are
larger for semidilute solutions, but the asymptotic Weissenberg
number dependence seems to be the same. At $c/c^*>1$ a universal
curve is adopted, as already pointed out above in connection with
Fig~\ref{P_phi_semidilute}.

%The increase of the width  $\Delta \varphi$ is consistent with the
%increase of the tumbling frequency, because a higher frequency
%implies a faster sampling of angles $\varphi$.

\begin{figure}[t]
%\vspace{0.6cm}
\begin{center}
\includegraphics*[width=.38\textwidth]{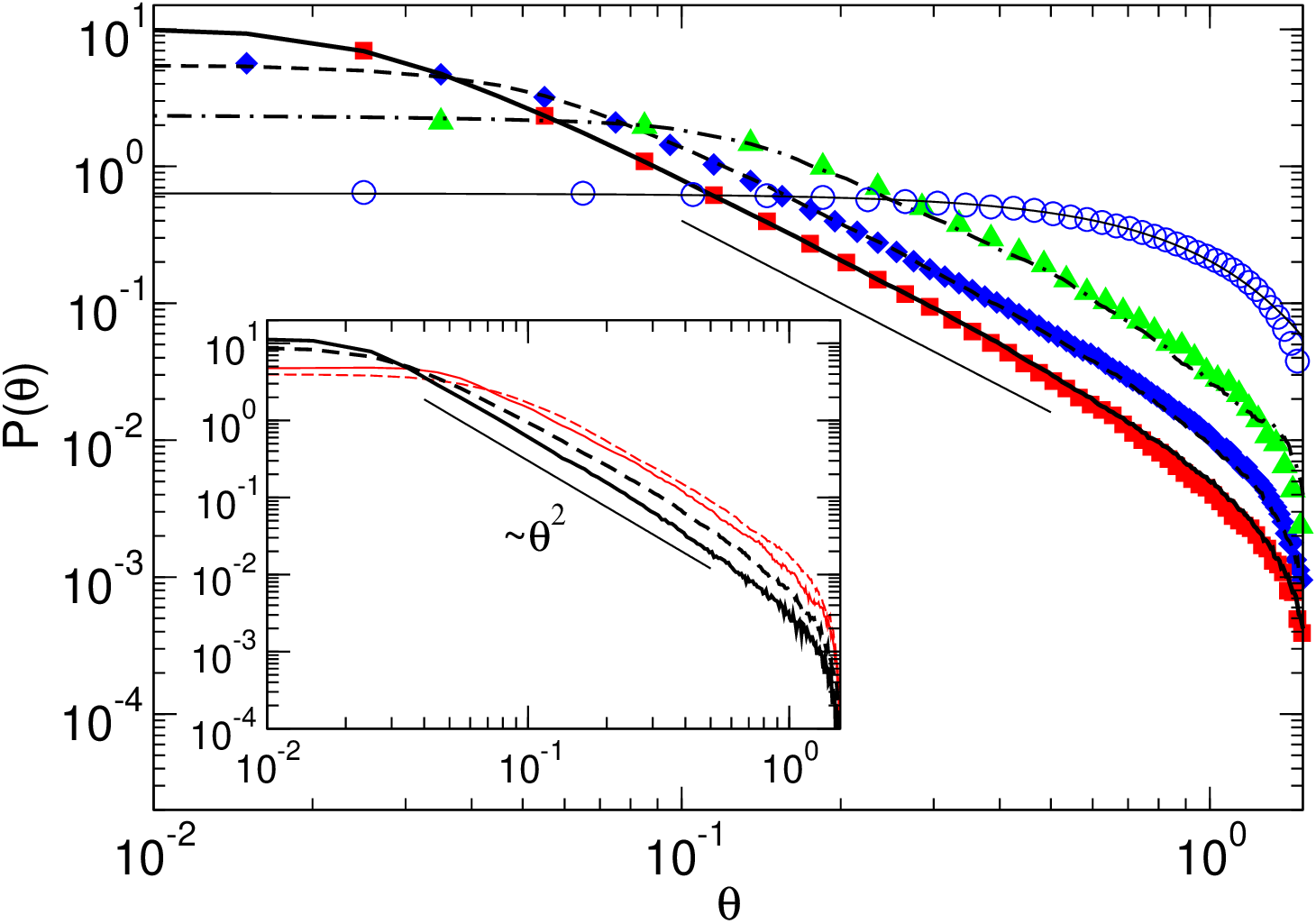}
\caption{Probability distribution functions $P(\theta)$ of polymers of
length $N_m = 250$ for
$c/c^*=2.77$ with $\wic=5520$ (${\color{red}\blacksquare}$),
$552$ (${\color{blue}\blacklozenge}$), $55.2$
(${\color{green}\blacktriangle}$), and $0.552$ ($\circ$),
as well as $c/c^*=10.38$ with $\wic=5691$ (solid line),
$569$ (dashed line), $57$ (dashed-dotted line), and $0.57$
(thin line).
The slope of the short solid lines is $-2$.
Inset: $P(\theta)$s of a dilute solution with $c/c^*=0.35$ for
$\wic=2670$ (black solid line) and $\wic=267$ (red solid line) and a
semidilute solution with $c/c^*=5.19$
for $\wic=2700$ (black dashed line) and $\wic=270$ (red dashed line).}
\label{theta_fig}
\end{center}
\end{figure}

Probability distribution functions of the angle $\theta$ are
displayed in Fig.~\ref{theta_fig} for various Weissenberg numbers.
Theoretical calculations for dilute solutions predict a crossover
from a Gaussian shape of the distribution function to a power-law
decay according to $P(\theta)\sim\theta^{-2}$ with increasing
shear rate, within a certain range of angles, which is confirmed
by the simulations. The dependence of $P(\theta)$ on concentration
is in accord with that of $P(\varphi)$. In the semidilute regime,
$P(\theta)$ is independent of concentration for a given $\wic$,
while a comparison of distributions of \textit{semidilute} and
\textit{dilute} solutions yields a broadening for the semidilute
case.

We attribute the concentration  independence of the tumbling time
and the probability distribution functions for $c/c^*> 1$ to
screening of hydrodynamic interactions. To confirm our hypothesis,
we performed Brownian MPC simulations for dilute and semidilute
solutions. Using similar Weissenberg numbers, we find, within the
accuracy of the simulations, identical distribution functions
$P(\varphi)$ for both cases. Moreover, the distributions agree
with those of semidilute systems of the same concentration and
Weissenberg number in the presence of hydrodynamic interactions.
Hence, the differences between distribution functions at low and
high concentrations, as displayed in the insets of Figs.~6 and 8,
are due to hydrodynamic interactions. In dilute solutions,
hydrodynamic interactions are present, whereas in systems with $c
> c^*$, hydrodynamic interactions are screened. Naturally, at
larger concentrations friction is higher. This aspect is captured
in the relaxation time $\tau(c)$, which increases considerably
with concentration \cite{huan:10_1}.

The fact that the tumbling frequencies and the widths of the
distribution functions are larger in semidilute solutions, i.e.,
when hydrodynamic interactions are screened, might be explained as
follows. (i) The observed broadening of the distribution function
$P(\varphi)$ in semidilute solution implies that hydrodynamic
interactions favor polymer alignment and lead to a faster dynamics
during the collapse and stretching part of the tumbling motion.
(ii) At the same Weissenberg number, the shear rate of a
non-draining polymer is larger than that of a free-draining one,
due to differences in equilibrium relaxation times, i.e., for a
coiled conformation. As a consequence, the effective Weissenberg
number $\wi_R= \dot \gamma \tau_R$, where $\tau_R$ is the
rotational relaxation time in the stretched rodlike conformation,
of the non-draining polymer is larger than that of the
free-draining one. This could explain the larger probability of
angles in the vicinity of $\varphi_m$ for non-draining polymers as
well as their faster collapse dynamics. Overall, the tumbling time
is larger in a non-draining system.

The broadening of the distribution functions with increasing
concentration or screening of hydrodynamic interactions is not
captured by standard theories employing the preaveraging
approximation \cite{doi:86,wink:06_1,wink:10}. Here, hydrodynamic
interactions are included in the relaxation times and hence the
Weissenberg number only. Additional, ``higher order effects'' are
neglected. Therefore, one might expect that the theoretical
description would reproduce results of simulations without
hydrodynamic interactions, in contrast, the model calculations
rather reproduce the simulation data for systems with hydrodynamic
interactions.

\section{V. Conclusions}

We have found in Ref. \cite{huan:10_1} that in shear flow the
stationary-state conformational polymer properties are independent
of concentration when expressed in terms of the Weissenberg number
$\wic = \dot \gamma \tau(c)$. This is remarkable, since the
longest polymer relaxation time $\tau(c)$ increases significantly
with concentration and indicates that an effective local friction
determines the stationary-state properties.

Here, we have analyzed dynamical properties---orientational
distribution functions and tumbling times---of semidilute polymer
solutions in shear flow and have found that they depend on
concentration (in excess of $\tau(c)$), a  dependence which we
attribute to screening of hydrodynamic interactions in semidilute
solution. Compared to the dilute case, such a screening causes a
broadening of orientational angle distribution functions and an
increasing ratio $f=\tau_t/\tau(c)$ at the same Weissenberg number
$\wic$ in semidilute solution. The effect itself is small
($f(c=0)/f(c>c^*) \approx 1.3$ at $\wic = 10^3$). More
importantly, the same asymptotic dependencies are obtained as
function of the Weissenberg number $\wic$  in dilute and
semidilute solutions. This explains the previously obtained
agreement of power spectral densities obtained from free-draining
and non-draining computer simulations \cite{schr:05}.

\acknowledgments
 Financial support by the Deutsche Forschungsgemeinschaft
 within SFB TR6 is gratefully acknowledged. We are grateful to the
 J\"ulich Supercomputer Centre (JSC) for allocation of a special
 CPU-time grant.

%\bibliographystyle{eplbib}
%\bibliography{polymer_channel}

\begin{thebibliography}{10}
\expandafter\ifx\csname
url\endcsname\relax\def\url#1{\texttt{#1}}\fi

\bibitem{smith:99}
\Name{Smith D.~E., Babcock H.~P. \and Chu S.} \REVIEW{Science
  }{283}{1999}{1724}.

\bibitem{schr:05}
\Name{Schroeder C.~M., Teixeira R.~E., Shaqfeh E. S.~G. \and Chu
S.}
  \REVIEW{Phys. Rev. Lett. }{95}{2005}{018301}.

\bibitem{teix:05}
\Name{Teixeira R.~E., Babcock H.~P., Shaqfeh E. S.~G. \and Chu S.}
  \REVIEW{Macromolecules }{38}{2005}{581}.

\bibitem{schr:05_1}
\Name{Schroeder C.~M., Teixeira R.~E., Shaqfeh E. S.~G. \and Chu
S.}
  \REVIEW{Macromolecules }{38}{2005}{1967}.

\bibitem{gera:06}
\Name{Gerashchenko S. \and Steinberg V.} \REVIEW{Phys. Rev. Lett.
  }{96}{2006}{038304}.

\bibitem{aust:99}
\Name{Aust C., Kr{\"o}ger M. \and Hess S.} \REVIEW{Macromolecules
  }{32}{1999}{5660}.

\bibitem{doyl:00}
\Name{Doyle P.~S., Ladoux B. \and Viovy J.-L.} \REVIEW{Phys. Rev.
Lett.
  }{84}{2000}{4769}.

\bibitem{cela:05}
\Name{Celani A., Puliafito A. \and Turitsyn K.} \REVIEW{Europhys.
Lett.
  }{70}{2005}{464}.

\bibitem{cher:05}
\Name{Chertkov M., Kolokolov I., Lebedev A. \and Turitsyn K.}
\REVIEW{J. Fluid.
  Mech. }{531}{2005}{251}.

\bibitem{puli:05}
\Name{Puliafito A. \and Turitsyn K.} \REVIEW{Physica D
}{211}{2005}{9}.

\bibitem{delg:06}
\Name{Delgado-Buscalioni R.} \REVIEW{Phys. Rev. Lett.
}{96}{2006}{088303}.

\bibitem{wink:06_1}
\Name{Winkler R.~G.} \REVIEW{Phys. Rev. Lett. }{97}{2006}{128301}.

\bibitem{zhan:09}
\Name{Zhang Y., Donev A., Weisgraber T., Alder B.~J., Graham M.~G.
\and
  de~Pablo J.~J.} \REVIEW{J. Chem. Phys. }{130}{2009}{234902}.

\bibitem{babc:00}
\Name{Babcock H.~P., Smith D.~E., Hur J.~S., Shaqfeh E. S.~G. \and
Chu S.}
  \REVIEW{Phys. Rev. Lett. }{85}{2000}{2018}.

\bibitem{hur:01}
\Name{Hur J., Shaqfeh E. S.~G., Babcock H.~P., Smith D.~E. \and
Chu S.}
  \REVIEW{J. Rheol. }{45}{2001}{421}.

\bibitem{huan:10_1}
\Name{Huang C.-C., Winkler R.~G., Sutmann G. \and Gompper G.}
  \REVIEW{Macromolecules }{43}{2010}{10107}.

\bibitem{doi:86}
\Name{Doi M. \and Edwards S.~F.} \Book{The Theory of Polymer
Dynamics}
  (Clarendon Press, Oxford) 1986.

\bibitem{kapr:08}
\Name{Kapral R.} \REVIEW{Adv. Chem. Phys. }{140}{2008}{89}.

\bibitem{gomp:09}
\Name{Gompper G., Ihle T., Kroll D.~M. \and Winkler R.~G.}
\REVIEW{Adv. Polym.
  Sci. }{221}{2009}{1}.

\bibitem{male:99}
\Name{Malevanets A. \and Kapral R.} \REVIEW{J. Chem. Phys.
}{110}{1999}{8605}.

\bibitem{galu:10}
\Name{Galuschko A., Spirin L., Kreer T., Johner A., Pastorino C.,
Wittmer J.
  \and Baschnagel J.} \REVIEW{Langmuir }{26}{2010}{6418}.

\bibitem{alle:87}
\Name{Allen M.~P. \and Tildesley D.~J.} \Book{Computer Simulation
of Liquids}
  (Clarendon Press, Oxford) 1987.

\bibitem{ihle:01}
\Name{Ihle T. \and Kroll D.~M.} \REVIEW{Phys. Rev. E
}{63}{2001}{020201(R)}.

\bibitem{huan:10}
\Name{Huang C.-C., Chatterji A., Sutmann G., Gompper G. \and
Winkler R.~G.}
  \REVIEW{J. Comput. Phys. }{229}{2010}{168}.

\bibitem{lamu:01}
\Name{Lamura A., Gompper G., Ihle T. \and Kroll D.~M.}
\REVIEW{Europhys. Lett.
  }{56}{2001}{319}.

\bibitem{wink:09}
\Name{Winkler R.~G. \and Huang C.-C.} \REVIEW{J. Chem. Phys.
  }{130}{2009}{074907}.

\bibitem{ripo:07}
\Name{Ripoll M., Winkler R.~G. \and Gompper G.} \REVIEW{Eur. Phys.
J. E
  }{23}{2007}{349}.

\bibitem{wink:10}
\Name{Winkler R.~G.} \REVIEW{J. Chem. Phys. }{133}{2010}{164905}.

\bibitem{jose:08}
\Name{Jose P.~P. \and Szamel G.} \REVIEW{J. Chem. Phys.
}{128}{2008}{224910}.

\end{thebibliography}

\end{document}